\documentclass[journal]{IEEEtran}
\usepackage{graphicx,amsmath,url}
\usepackage{microtype,lmodern}
\usepackage{mathtools,amssymb,gensymb}
\usepackage{tgtermes} 
\usepackage[T1]{fontenc}
\usepackage[utf8]{inputenc}
\usepackage{color,soul}
\usepackage[colorinlistoftodos]{todonotes}
\usepackage{acronym}
\usepackage{array}
\usepackage{booktabs}
\usepackage[compress]{cite}

\usepackage[normalem]{ulem}
\usepackage[switch]{lineno}
\usepackage{listings}
 
\usepackage{longtable}
\usepackage[colorlinks,breaklinks]{hyperref} 

\hypersetup{hypertexnames=true,
  linkcolor=blue,anchorcolor=black,citecolor=blue,urlcolor=blue
}
\graphicspath{{imglocal/}{img/}}
\acrodefplural{RAM}[RAMs]{Random Access Memories}
\acrodefplural{RRAM}[R-RAMs]{Resistive Random Access Memories}
\acrodefplural{STT-MRAM}[STT-MRAMs]{Spin-Transfer Torque Magnetic Random Access Memories}
\acrodef{ADC}[ADC]{Analog to Digital Converter}
\acrodef{ADEX}[AdExp-I\&F]{Adaptive-Exponential Integrate and Fire}
\acrodef{AER}[AER]{Address-Event Representation}
\acrodef{AEX}[AEX]{AER EXtension board}
\acrodef{AE}[AE]{Address-Event}
\acrodef{AFM}[AFM]{Atomic Force Microscope}
\acrodef{AGC}[AGC]{Automatic Gain Control}
\acrodef{AMDA}[AMDA]{AER Motherboard with D/A converters}
\acrodef{ANN}[ANN]{Attractor Neural Network}
\acrodef{API}[API]{Application Programming Interface}
\acrodef{ARM}[ARM]{Advanced RISC Machine}
\acrodef{ASIC}[ASIC]{Application Specific Integrated Circuit}
\acrodef{BCM}[BMC]{Bienenstock-Cooper-Munro}
\acrodef{BD}[BD]{Bundled Data}
\acrodef{BEOL}[BEOL]{Back-end of Line}
\acrodef{BG}[BG]{Bias Generator}
\acrodef{BMI}[BMI]{Brain-Machince Interface}
\acrodef{CAD}[CAD]{Computer Aided Design}
\acrodef{CAM}[CAM]{Content Addressable Memory}
\acrodef{CAVIAR}[CAVIAR]{Convolution AER Vision Architecture for Real-Time}
\acrodef{CFC}[CFC]{Current to Frequency Converter}
\acrodef{CCN}[CCN]{Cooperative and Competitive Network}
\acrodef{CHP}[CHP]{Communicating Hardware Processes}
\acrodef{CNN}[CCN]{Convolutional Neural Network}
\acrodef{CMIM}[CMIM]{Metal-insulator-metal Capacitor}
\acrodef{CMOL}[CMOL]{``Hybrid CMOS nanoelectronic circuits''}
\acrodef{CMOS}[CMOS]{Complementary Metal-Oxide-Semiconductor}
\acrodef{COTS}[COTS]{Commercial Off-The-Shelf}
\acrodef{CPG}[CPG]{Central Pattern Generator}
\acrodef{CPLD}[CPLD]{Complex Programmable Logic Device}
\acrodef{CPU}[CPU]{Central Processing Unit}
\acrodef{CV}[CV]{Coefficient of Variation}
\acrodef{DAC}[DAC]{Digital to Analog Converter}
\acrodef{DAS}[DAS]{Dynamic Auditory Sensor}
\acrodef{DAVIS}[DAVIS]{Dynamic and Active Pixel Vision Sensor}
\acrodef{DBN}[DBN]{Deep Belief Network}
\acrodef{DFA}[DFA]{Deterministic Finite Automaton}
\acrodef{DMA}[DMA]{Direct Memory Access}
\acrodef{DNF}[DNF]{Dynamic Neural Field}
\acrodef{DNN}[DNN]{Deep Neural Network}
\acrodef{DOF}[DOF]{Degrees of Freedom}
\acrodef{DPE}[DPE]{Dynamic Parameter Estimation}
\acrodef{DPI}[DPI]{Differential Pair Integrator}
\acrodef{DRAM}[DRAM]{Dynamic Random Access Memory}
\acrodef{DR}[DR]{Dual Rail}
\acrodef{DSP}[DSP]{Digital Signal Processor}
\acrodef{DVS}[DVS]{Dynamic Vision Sensor}
\acrodef{EBL}[EBL]{Electron Beam Lithography}
\acrodef{EDVAC}[EDVAC]{Electronic Discrete Variable Automatic Computer}
\acrodef{EIN}[EIN]{Excitatory-Inhibitory Network}
\acrodef{EM}[EM]{Expectation Maximization}
\acrodef{EPSC}[EPSC]{Excitatory Post-Synaptic Current}
\acrodef{EPSP}[EPSP]{Excitatory Post-Synaptic Potential}
\acrodef{FDSOI}[FDSOI]{Fully-Depleted Silicon on Insulator}
\acrodef{FET}[FET]{Field-Effect Transistor}
\acrodef{FFT}[FFT]{Fast Fourier Transform}
\acrodef{FI}[F-I]{Frequency-Current}
\acrodef{FPGA}[FPGA]{Field Programmable Gate Array}
\acrodef{FSA}[FSA]{Finite State Automaton}
\acrodef{FSM}[FSM]{Finite State Machine}
\acrodef{GOPS}[GOPS]{Giga-Operations per Second}
\acrodef{GPU}[GPU]{Graphical Processing Unit}
\acrodef{GUI}[GUI]{Graphical User Interface}
\acrodef{HAL}[HAL]{Hardware Abstraction Layer}
\acrodef{HH}[H\&H]{Hodgkin \& Huxley}
\acrodef{HMM}[HMM]{Hidden Markov Model}
\acrodef{HRS}[HRS]{High-Resistive State}
\acrodef{HR}[HR]{Human Readable}
\acrodef{HSE}[HSE]{Handshaking Expansion}
\acrodef{HW}[HW]{Hardware}
\acrodef{ICT}[ICT]{Information and Communication Technology}
\acrodef{IC}[IC]{Integrated Circuit}
\acrodef{IF2DWTA}[IF2DWTA]{Integrate \& Fire 2--Dimensional WTA}
\acrodef{IFSLWTA}[IFSLWTA]{Integrate \& Fire Stop Learning WTA}
\acrodef{IF}[I\&F]{Integrate-and-Fire}
\acrodef{IMU}[IMU]{Inertial Measurement Unit}
\acrodef{INCF}[INCF]{International Neuroinformatics Coordinating Facility}
\acrodef{INI}[INI]{Institute of Neuroinformatics}
\acrodef{IO}[I/O]{Input/Output}
\acrodef{IPSC}[IPSC]{Inhibitory Post-Synaptic Current}
\acrodef{IPSP}[IPSP]{Inhibitory Post-Synaptic Potential}
\acrodef{IP}[IP]{Intellectual Property}
\acrodef{ISI}[ISI]{Inter-Spike Interval}
\acrodef{JFLAP}[JFLAP]{Java - Formal Languages and Automata Package}
\acrodef{LLC}[LLC]{Low Leakage Cell}
\acrodef{LFP}[LFP]{Local Field Potential}
\acrodef{LNA}[LNA]{Low-Noise Amplifier}
\acrodef{LPF}[LPF]{Low-Pass Filter}
\acrodef{LRS}[LRS]{Low-Resistive State}
\acrodef{LSM}[LSM]{Liquid State Machine}
\acrodef{LTD}[LTD]{Long Term Depression}
\acrodef{LTI}[LTI]{Linear Time-Invariant}
\acrodef{LTP}[LTP]{Long Term Potentiation}
\acrodef{LTU}[LTU]{Linear Threshold Unit}
\acrodef{LUT}[LUT]{Look-Up Table}
\acrodef{MCMC}[MCMC]{Markov-Chain Monte Carlo}
\acrodef{MEMS}[MEMS]{Micro Electro Mechanical System}
\acrodef{MIM}[MIM]{Metal Insulator Metal}
\acrodef{MOSCAP}[MOSCAP]{Metal Oxide Semiconductor Capacitor}
\acrodef{MOSFET}[MOSFET]{Metal Oxide Semiconductor Field-Effect Transistor}
\acrodef{MOS}[MOS]{Metal Oxide Semiconductor}
\acrodef{MRI}[MRI]{Magnetic Resonance Imaging}
\acrodef{NDFSM}[NDFSM]{Non-deterministic Finite State Machine} 
\acrodef{ND}[ND]{Noise-Driven}
\acrodef{NEF}[NEF]{Neural Engineering Framework}
\acrodef{NHML}[NHML]{Neuromorphic Hardware Mark-up Language}
\acrodef{NIL}[NIL]{Nano-Imprint Lithography}
\acrodef{NMDA}[NMDA]{N-Methyl-D-Aspartate}
\acrodef{NME}[NE]{Neuromorphic Engineering}
\acrodef{OTA}[OTA]{Operational Transconductance Amplifier}
\acrodef{PCB}[PCB]{Printed Circuit Board}
\acrodef{PFM}[PFM]{Pulse Frequency Modulation}
\acrodef{PR}[PR]{Production Rule}
\acrodef{PSC}[PSC]{Post-Synaptic Current}
\acrodef{PSTH}[PSTH]{Peri-Stimulus Time Histogram}
\acrodef{QDI}[QDI]{Quasi Delay Insensitive}
\acrodef{RAM}[RAM]{Random Access Memory}
\acrodef{RMSE}[RMSE]{Root Mean Squared-Error}
\acrodef{RMS}[RMS]{Root Mean Squared}
\acrodef{RNN}[RNN]{Recurrent Neural Network}
\acrodef{ROLLS}[ROLLS]{Reconfigurable On-Line Learning Spiking}
\acrodef{RRAM}[RRAM]{Resistive Random Access Memory}
\acrodef{SAC}[SAC]{Selective Attention Chip}
\acrodef{SCX}[SCX]{Silicon CorteX}
\acrodef{SD}[SD]{Signal-Driven}
\acrodef{SEM}[SEM]{Spike-based Expectation Maximization}
\acrodef{SLAM}[SLAM]{Simultaneous Localization and Mapping}
\acrodef{SOC}[SOC]{System-On-Chip}
\acrodef{SOI}[SOI]{Silicon on Insulator}
\acrodef{SRAM}[SRAM]{Static Random Access Memory}
\acrodef{STDP}[STDP]{Spike-Timing Dependent Plasticity}
\acrodef{STD}[STD]{Short-Term Depression}
\acrodef{STP}[STP]{Short-Term Plasticity}
\acrodef{STT-MRAM}[STT-MRAM]{Spin-Transfer Torque Magnetic Random Access Memory}
\acrodef{STT}[STT]{Spin-Transfer Torque}
\acrodef{SW}[SW]{Software}
\acrodef{TFT}[TFT]{Thin Film Transistor}
\acrodef{USB}[USB]{Universal Serial Bus}
\acrodef{VHDL}[VHDL]{VHSIC Hardware Description Language}
\acrodef{VLSI}[VLSI]{Very Large Scale Integration}
\acrodef{VOR}[VOR]{Vestibulo-Ocular Reflex}
\acrodef{WTA}[WTA]{Winner-Take-All}
\acrodef{XML}[XML]{eXtensible Mark-up Language}
\acrodef{divmod3}[DIVMOD3]{divisibility of a number by 3}
\acrodef{hWTA}[hWTA]{Hard Winner-Take-All}
\acrodef{sWTA}[sWTA]{soft Winner-Take-All}

\begin{document}
%

\title{An auto-scaling wide dynamic range current to frequency converter for real-time monitoring of signals in neuromorphic systems}

\author{\IEEEauthorblockN{Ning Qiao and Giacomo~Indiveri}
\IEEEauthorblockA{Institute of Neuroinformatics, University of Zurich and ETH Zurich\\ 
Zurich, Switzerland\\
Email: [qiaoning$|$giacomo]@ini.uzh.ch}}


%


\maketitle

\begin{abstract}
Neuromorphic systems typically employ current-mode circuits that model neural dynamics and produce output currents that range from few pico-Amperes to hundreds of micro-Amperes. On-line real-time monitoring of the signals produced by these circuits is crucial, for prototyping and debugging purposes, as well as for analyzing and understanding the network dynamics and computational properties.  To this end, we propose a compact on-chip auto-scaling \ac{CFC} for real-time monitoring of analog currents in mixed-signal/analog neuromorphic electronic systems. The proposed \ac{CFC} is a self-timed asynchronous circuit that has a wide dynamic input range of up to 6 decades, ranging from pico-Amps to micro-Amps, with high current measurement sensitivity. To produce a linear output frequency response, while properly covering the wide dynamic input range, the circuit automatically detects the scale of the input current and adjusts the scale of its output firing rate accordingly.
Here we describe the proposed circuit and present experimental results measured from multiple instances of the circuit, implemented using a standard 180\,nm CMOS process, and interfaced to silicon neuron and synapse circuits for real-time current monitoring. 
We demonstrate how the circuit is suitable for measuring neural dynamics by showing the converted response properties of the chip silicon neurons and synapses as they are stimulated by input spikes.  \end{abstract}


%
\IEEEpeerreviewmaketitle

\acresetall

\section{Introduction}
Mixed-signal analog/digital neuromorphic \ac{VLSI} systems often employ current-mode circuits to emulate the properties of biological neural systems, and to implement their computational principles~\cite{Mead90}. By exploiting their transistor's subtreshold region of operation and by interfacing them to asynchronous \ac{AER} digital communication modules~\cite{Boahen00}, these analog circuits are suitable for implementing complex and large-scale spiking neural networks in very low-power and compact dedicated \ac{VLSI} systems~\cite{Benjamin_etal14,Chicca_etal14}. During the system prototyping phases, basic research experiments, and neuromorphic system deployment in practical applications it is very useful and often necessary to perform on-line monitoring of the system's internal nodes and state variables at run-time. While this is something very costly in terms of memory and I/O bandwidth resources with digital neuromorphic systems, it is relatively simple to implement, in mixed signal analog/digital neuromorphic systems. However, this is true only for monitoring the \emph{voltages} of internal nodes, and not currents. For monitoring analog voltage signals it is sufficient to use digital circuits to select the nodes that should be monitored and high precision voltage buffers for sending the signals to pads and off-chip. However log-domain current-mode circuits present additional challenges: while it is possible in principle to monitor voltages converted from current-mode circuits using the same buffers, in practice this is not desirable, because the conversion typically involves a compressing logarithmic non-linearity. For this reason it is preferable to monitor directly the currents of such circuits, especially if they represent dynamic variables that emulate biophysically realistic temporal properties of real neural circuits. Existing approaches for monitoring currents are mainly based on precise clocked techniques~\cite{Linares-Barranco_Serrano-Gotarredona03,Voulgari_etal15} which are not suitable for continuous time log-domain circuits embedded in asynchronous event-driven systems.
In this paper we propose a novel asynchronous auto-scaling \ac{CFC}, based on a low-power programmable current integrator design, that is optimally suited for current-mode neuromorphic circuits, and which linearly converts the monitored current to spike rates, via \ac{PFM}.


In the next Section, we describe the circuits that implement the proposed \ac{CFC}. In Section~\ref{sec:experimental-results} we present experimental results obtained from the measurements of test circuits fabricated using a standard 180\,nm CMOS process, and in Section~\ref{sec:conclusions} we present concluding remarks that outline the benefits and potential of the approach proposed.


\begin{figure}
  \centering
  \includegraphics[width=0.4\textwidth]{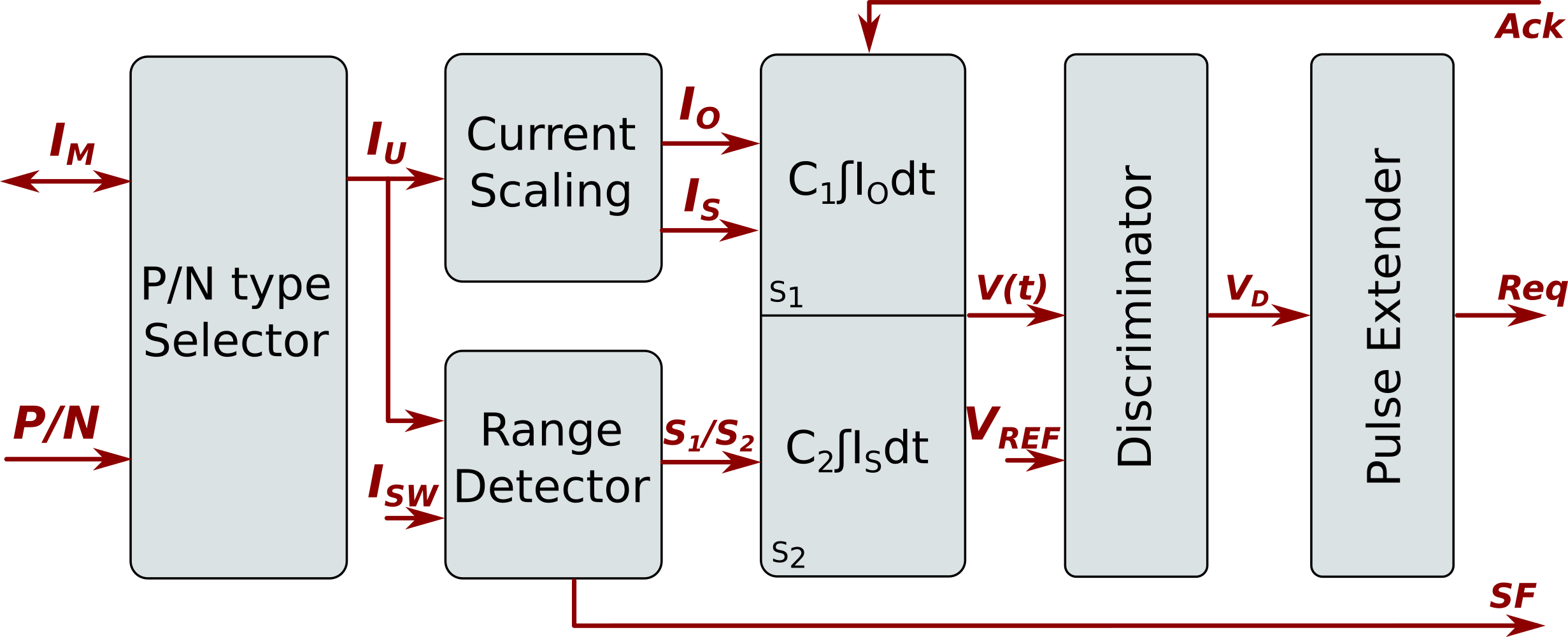}
  \caption{Block diagram describing the architecture of proposed \ac{CFC}. The monitored current passes first through a \emph{P/N type Selector} block; its rectified version is then copied to a \emph{Current Scaling} block, while in parallel its value is evaluated by a \emph{Range Detector} block; the scaled current from \emph{Current Scaling} block is then integrated using a capacitor that is selected by the \emph{Range Detector} block output signal; the \emph{Discriminator} and \emph{Pulse Extender} blocks produce Pulse-Frequency-Modulated voltage pulses using the \ac{AER} protocol, with a rate that is linearly proportional to the integrated current signal.}
  \label{fig:cfc_arch}
\end{figure}

\section{Current to Frequency Converter Circuit}
\label{sec:circuit-architecture}

\begin{figure}
  \centering
  \includegraphics[width=0.475\textwidth]{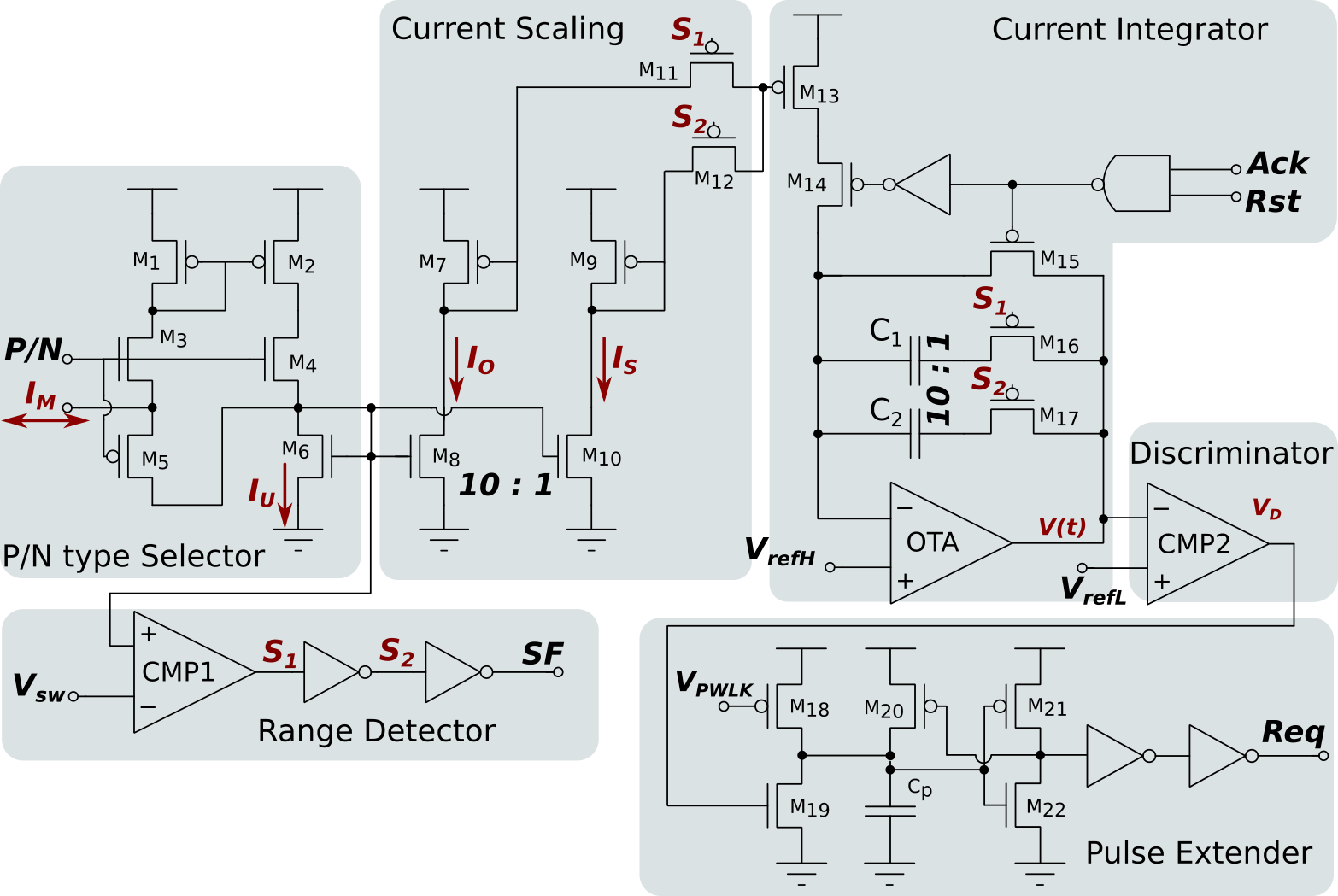}
  \caption{Circuit schematic of the proposed \ac{CFC}. The monitored current $I_{M}$ is rectified as $I_{U}$, mirrored as $I_{O}$, and scaled down as $I_{S}$. In parallel, the \emph{Range Detector} circuit evaluates $I_{M}$ and produces the signals $S_{1}$ and $S_{2}$ for integrating either $I_{O}$ on $C_{1}$ or $I_{S}$ on $C_{2}$ by transistors $M_{11,12,16,17}$. The voltage $V(t)$ gradually decreases during the integration phase, from it's initial value $V_{refH}$; once $V(t)$ reaches $V_{refL}$, it triggers a $Req$ pulse, of extended duration, regulated by the \emph{Pulse Extender} block. The integrator is reset once it receives an acknowledge signal $Ack$, from an \ac{AER} compatible receiver module.} 
  \label{fig:cfc_sch}
\end{figure}

Neuromorphic systems that employ current-mode sub-threshold circuits to model neural dynamics usually produce currents that range from few pico-Amperes to tens of micro-Amperes, with time constants that range from tens to hundreds of milliseconds. To accurately reproduce these signals, the \ac{CFC} needs to have a wide enough input dynamic range, and a high enough time resolution. Indeed, the \ac{CFC} circuit should produce spikes at rates that are high enough to allow an accurate reconstruction of the  monitored transient current. To convert linearly currents that span six orders of magnitude e.g., ranging from 1\,pA to 1\,$\mu$A, using for example output frequencies that range from 10\,Hz to 10M\,Hz, it would be necessary to allocate a significant amount of resources, in terms of area and power consumption. For this reason, we propose a design that senses the scale of the input current and re-scales it, automatically adapting the output frequency and keeping it within a limited manageable dynamic range.
The block diagram of the proposed \ac{CFC} design is shown in Fig.~\ref{fig:cfc_arch}. The circuit consists of a \emph{P/N type Selector} block for rectifying the input current; a \emph{Range Detector} and a \emph{Current Scaling} block for sensing and scaling the rectified current; a \emph{Current Integrator} block which chooses one of two options for integrating the scaled current; and a \emph{Discriminator} block which determines when to trigger an output pulse, whose duration is controlled by the \emph{Pulse Extender} block.
The detailed circuit schematics of the \ac{CFC} circuit are shown in Fig.~\ref{fig:cfc_sch}.

The \emph{P/N type selector} circuits rectifies the input current $I_{M}$, via the digitally controlled transistors $M_{3}-M_{5}$ to produce a current $I_{U}$ which is copied by $M_{6}$-$M_{8}$ to the $I_{O}$ current and by $M_{6}$-$M_{10}$ to the $I_{S}$ current. Transistors $M_{8}$, $M_{10}$ are sized such that  $I_{S} = \beta I_{O}$, with $\beta < 1$. In parallel, the gate voltage of $M_{6}$ is compared to a reference voltage $V_{SW}$, by the \emph{Range Detector} block. Depending on the outcome of this comparison, two active-low digital signals $S_{1}$ and $S_{2}$ are generated, which encode either the condition $I_{M}<I_{SW}$ or $I_{M}>I_{SW}$, respectively. A third digital output voltage $SF$ is used to signal to the user the range that is currently being considered. Based on the output of this block, either the current $I_{O}$  or $I_{S}$ is copied, via $M_{7}$-$M_{13}$ or $M_{9}$-$M_{13}$, into the \emph{Current Integrator} block.
The \emph{Current Integrator} block comprises two integrating capacitors $C_{1}$ and $C_{2}$ with a ratio $C_{2}/C_{1} = \alpha$ and $\alpha > 1$. The same $S_{1}$ and $S_{2}$ signals generated by the \emph{Range Detector} block determine which of the two capacitors to use, for the current integration, via the transistors $M_{16}$ and $M_{17}$. Therefore, small currents get integrated with an integration factor of 1, while  large currents get integrated with a much smaller integration factor of $\beta /\alpha$. To initialize the integrator block it is possible to use the global reset signal $Rst$. At the beginning, both plates of the capacitors will be reset to $V_{refH}$ with $V(t=0) = V_{refH}$ as initial condition. Once the $Rst$ is de-asserted, the current through $M_{13}$ gets integrated on the selected capacitor, and the circuit's output voltage $V_{t}$ decreases accordingly. As soon as $V_{t}$ reaches the reference voltage $V_{refL}$, the output of the \emph{Discriminator} block $V_{D}$ will go high, and trigger the $Req$ to go high as well, for a duration that is set by $V_{PWLK}$ in the \emph{Pulse Extender} block. The $Req$  signal is sent to \ac{AER} interfacing circuits which, after encoding the \ac{CFC} channel address, transmitting the Address-Event to the off-chip receiver, and receiving the handshaking $Ack$ signal back, will reset the integrator. The \emph{Pulse Extender} block has been implemented to make sure that the reset pulse is long enough to fully reset the current integrator.

We can derive the inter-spike interval $\delta T$ that the \ac{CFC} circuit described above produces, in response to an input current $I_{mon}$:
\begin{equation}
  \label{eq:deltaT}
\delta T= \frac{\beta C(V_{refH}-V_{refL})}{\alpha I_{mon}}
\end{equation}
In our implementation we set the current mirror ratio $\beta = M_{10}/M_{9}$ to 10, and the integration capacitor ratio $\alpha = C_{2}/C_{1}$ to 10.
So for currents larger than a user-programmable reference value, the total scaling factor is  $\beta / \alpha = 100$.
The reference voltages $V_{refH}$ and $V_{refL}$ are also programmable biases, which can be further tuned for selecting a desired output firing rate. For example, if we set $V_{refH} - V_{refL} = 1V$,  consider capacitors $C_{1} = 100\,fF$ and $C_{2} = 1\,pF$, and set the scaling reference threshold such that currents $I_{mon} < 10\,nA$ are not scaled, and $I_{mon} > 10\,nA$ are scaled, then input current ranges from 1\,pA to 10\,nA will produce output firing rates that range from 10\,Hz to 100k\,Hz, and current ranges from 10\,nA to 1\,$\mu$A will produce output rates from 1k\,Hz to 100k\,Hz.
The response of the circuit is linear, provided that the duration of the reset pulse, extended by the \emph{Pulse Extender} block, is negligible with respect to the minimum inter-spike interval produced by the \ac{CFC}. For the capacitors sizes implemented in our test chip $C_{1} = 100\,fF$ and $C_{2} = 1\,pF$, reset pulse lengths of $T_{RST} = 0.1\,\mu$m are sufficiently smaller than the typical $\delta T$s produced.


\begin{figure}
  \centering
  \includegraphics[width=0.48\textwidth]{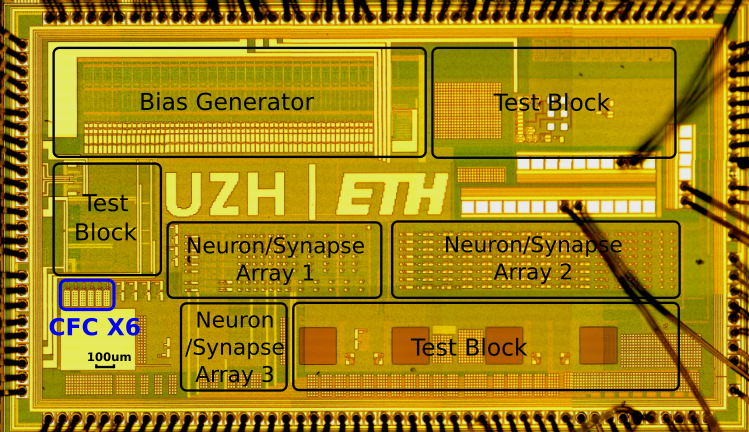}
  \caption{Die micro-photograph of the test chip with 6 \ac{CFC} channels (circled in blue) interfaced to different neuromorphic neuron/synapse arrays, for real-time monitoring of neural dynamics.}
  \label{fig:pioneer}
\end{figure}

\section{Experimental results}
\label{sec:experimental-results}

\begin{figure}
  \centering
  \includegraphics[width=0.47\textwidth]{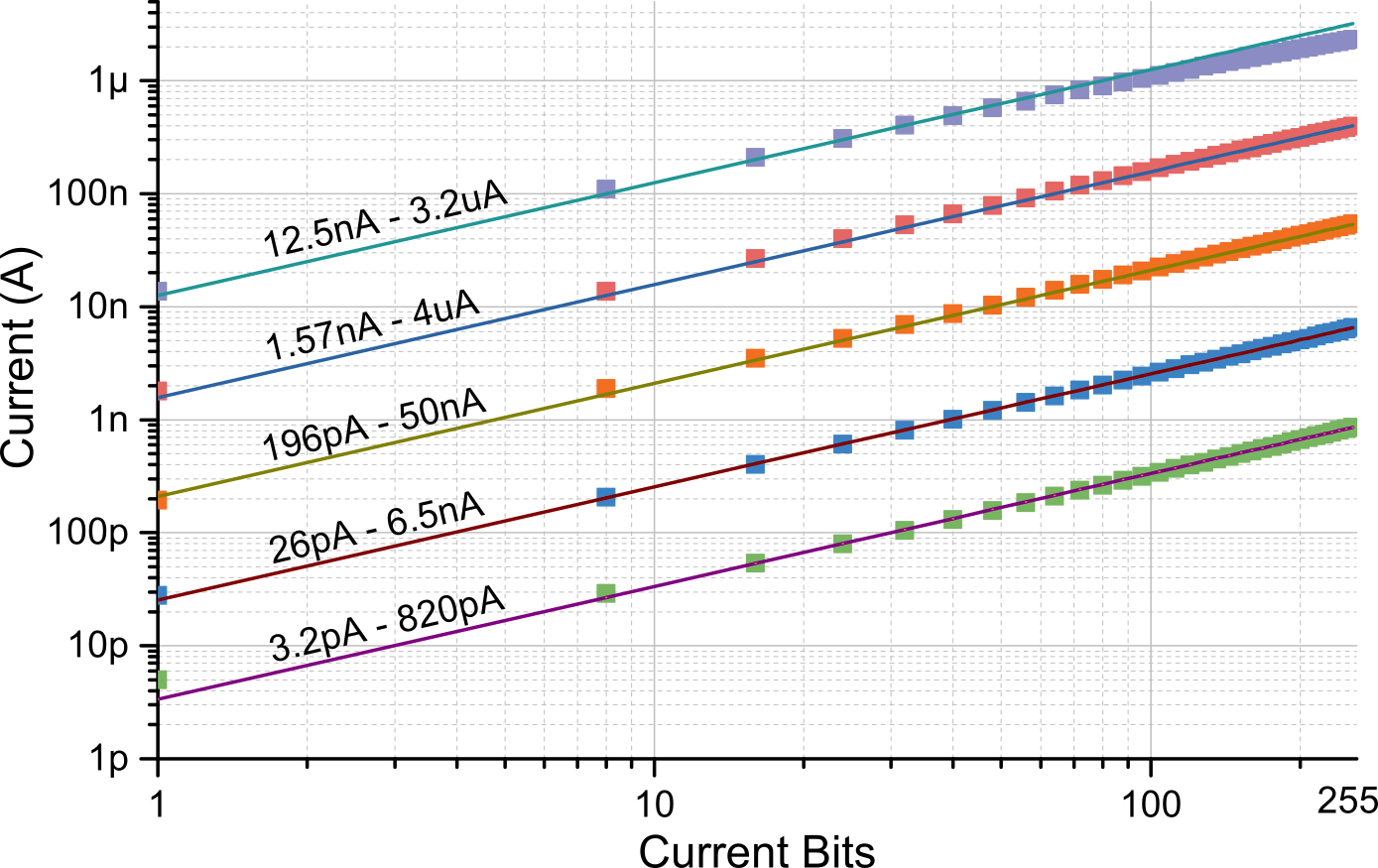}
  \caption{\ac{CFC} response (color symbols) to five linear current sweeps ( 3.2\,pA to 820\,pA, 26\,pA to 6.5\,nA, 196\,pA to 50\,nA, 1.57\,nA to 4\,$\mu$A, 12.5\,nA to 3.2\,$\mu$A) generated by p-FET transistors biased via an on-chip bias generator (color lines).}
  \label{fig:BG}
\end{figure}

We implemented six instances of the \ac{CFC} circuit described in Fig.~\ref{fig:cfc_sch} in a prototype test chip, fabricated using a standard 180\,nm CMOS process. The full chip occupies an area of 3.96\,mm$\times$2.29\,mm, and each \ac{CFC} block occupies an area of 150\,$\mu$m$\times$40\,$\mu$m (see Fig.~\ref{fig:pioneer} for chip micro-graph). We interfaced each of the \ac{CFC} blocks to multiple silicon neuron and synapse arrays for real-time current monitoring. In addition, we connected one of the converters to one p-FET transistor, with its source terminal tied to Vdd and drain terminal connected to the converter's input node. We used an on-chip programmable bias generator~\cite{Delbruck_etal10} to sweep the gate bias voltage of this p-FET transistor and to inject increasing current into the \ac{CFC}. We programme bias generator to span five different ranges of currents: 3.2\,pA to 820\,pA; 26\,pA to 6.5\,nA; 196\,pA to 50\,nA; 1.57\,nA to 4\,$\mu$A; and 12.5\,nA to 3.2\,$\mu$A. For each current branch, we swept the input current linearly by appropriately configuring the bias generator digital bits, and calculated the corresponding measured current from the \ac{CFC} outputs. Figure~\ref{fig:BG} shows the outcome of these measurements. As shown our circuit can accurately measure currents ranging from approximately 10\,pA to 1\,$\mu$A. Currents smaller than  5.5\,pA produce no effective output, mainly because of leakage issues in the current mirrors we used to copy current from one block to the next. On the other extreme, currents larger than 1\,$\mu$A lead to distortions, because of the non-linear effects introduced by the finite pulse width of the spikes $T_{RST}$, which start to become comparable to the $\delta T$ values produced by the \ac{CFC}.

\begin{figure}
  \centering
  \includegraphics[width=0.47\textwidth]{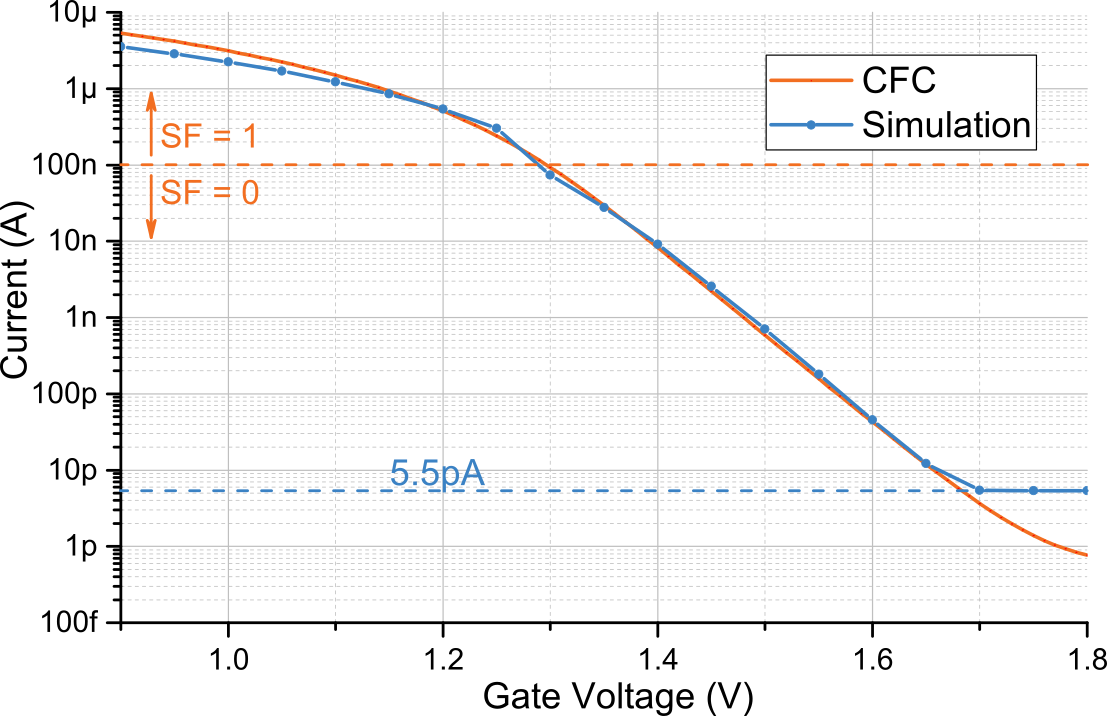}
  \caption{\ac{CFC} response to the current produced by a single p-MOS transistor while sweeping its gate voltage (orange line) compared to the transistor's corresponding SPICE simulation output (blue dot line).}
  \label{fig:PMOS}
\end{figure}

In Fig.~\ref{fig:PMOS} we compare the response of the \ac{CFC} circuit to the current produced by a single p-FET transistor while linearly sweeping its gate voltage with the current of that p-FET transistor derived from a corresponding SPICE simulation.
As  shown, the current derived from the  \ac{CFC} measurements matches the SPICE simulation outcome remarkably well for a wide range of currents, and distorts for currents smaller than 5.5\,pA and larger than 1$\mu$A. In this experiment, the scaling reference threshold threshold was set to 100\,nA.

\begin{figure}
  \centering
  \includegraphics[width=0.48\textwidth]{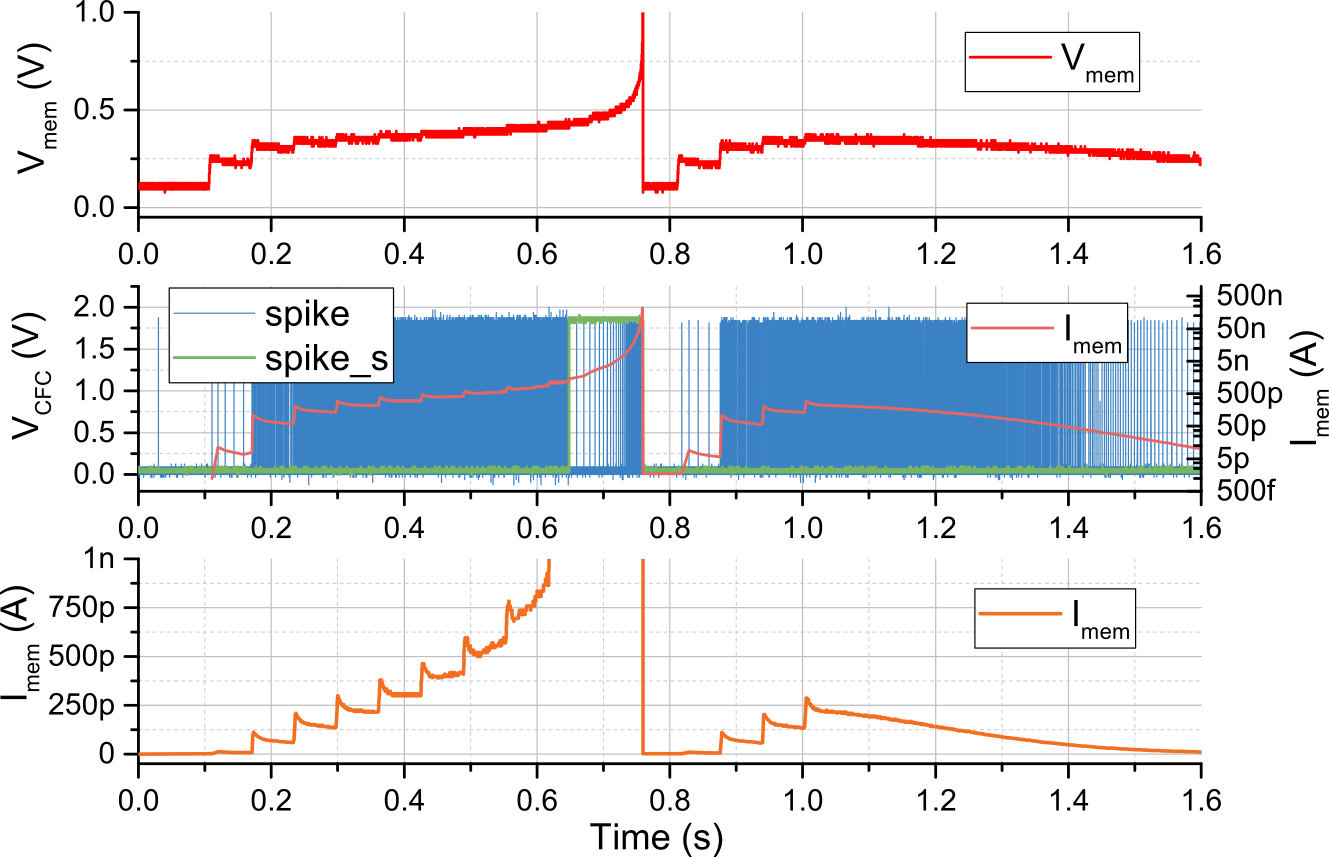}
  \caption{Neural dynamics in monitored silicon exponential \ac{IF} neuron. (Top) neuron's membrane potential from voltage buffer; (Medium) response of \ac{CFC} and represented membrane current in log-scale; (Bottom) zoom-in view of represented membrane current.}
  \label{fig:Imem}
\end{figure}

To provide examples of the dynamic  performance of the proposed \ac{CFC}, we provide measurements of neuromorphic current-mode neuron and synapse circuits~\cite{Chicca_etal14}, as they are being stimulated by input spikes and produce representative synapse and neural dynamics.
In Fig.~\ref{fig:Imem} we show the measurements from an \ac{ADEX} neuron circuit, as it is being stimulated via an excitatory synapse, with an input spike train of 20\,Hz. The top plot of  Fig.~\ref{fig:Imem} shows the voltage measured from the neuron's integrating capacitor, monitored by a high-gain voltage buffer. Given the current-mode nature of the circuit~\cite{Chicca_etal14}, this voltage actually represents the log() of the membrane potential variable. To best measure the dynamics of the variable equivalent to the true membrane potential in this circuit it is necessary to monitor the relevant current, and not voltage. In the second plot of Fig.~\ref{fig:Imem} the current (orange line) is derived from the pulses produced by the \ac{CFC} (blue pulses); the auto-range detection signal, shown in green, switches shortly after $t=600$\,ms, due to the large range spanned by the current. To plot the measured current in its full range, we used a semi-log scale (see right axes on figure). In this representation, the measured current resembles very much the measured voltage of the top plot (as expected). On the other hand, in the bottom plot of Fig.~\ref{fig:Imem}, we clip the range of the measured current, and plot it on a linear scale, highlighting the neuron's subthreshold dynamics. As shown, it is now evident that the log-domain neuron circuit exhibits an exponentially decaying conductance-based behavior in response to each spike, and a rising exponential dynamics as the current approaches the neuron's firing threshold, faithfully following  the behavior of the \ac{ADEX} computational model~\cite{Brette_Gerstner05}. 

\begin{figure}
  \centering
  \includegraphics[width=0.48\textwidth]{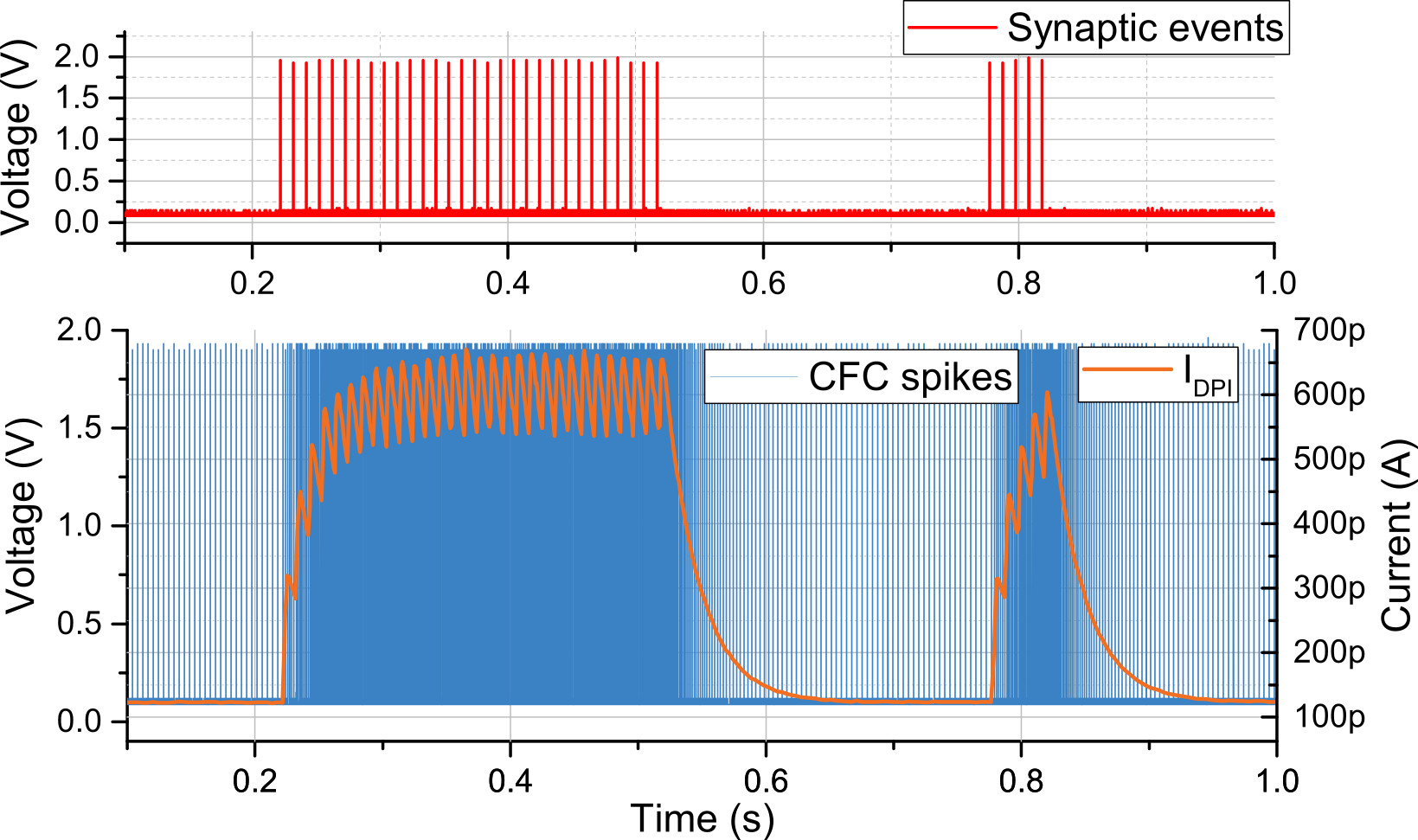}
  \caption{\ac{DPI} synapse current dynamics in response to input voltage spikes. (Top) input spike train. (Bottom) derived current measured from the \ac{CFC} pulses.}
  \label{fig:DPI}
\end{figure}

In Fig.~\ref{fig:DPI} we show measurements taken from a neuromorphic synapse circuit implemented using a log-domain \ac{DPI} circuit~\cite{Chicca_etal14}, in response to input voltage spikes. Using these measurements it is straightforward to derive the synapse model parameters, such as time-constant, and synaptic weight.

The power consumption of this circuit depends on its output rate. Thanks to automatic input range and output rate scaling features, spike rates are kept to low values even for very large input currents. In the worst-case scenario, for an output rate of 100k\,Hz (i.e., the maximum foreseen by this design), the power consumption is approximately 36\,nW.

\section{Conclusions}
\label{sec:conclusions}

We proposed a novel mixed-signal analog/digital auto-scaling current to frequency converter circuit design for measuring analog currents in neuromorphic electronic systems. The circuit is compact low-power and has a remarkable wide-input range. We validated the design in a 180\,nm CMOS process, demonstrated that it has an input dynamic range of up to 6 decades with high current measurement sensitivity, and showed how it is optimally suited for monitoring in real-time the signals produced by log-domain neuromorphic circuits. The compact and low-power nature of this circuit makes it ideal for being integrated in neuromorphic system that interact with the environment in real-time, and for building closed-loop control and adaptive set-ups accordingly.


\section*{Acknowledgments}

This work is supported by the EU ERC 
grant ``NeuroP'' (257219) and by the EU ICT grant ``NeuRAM$^3$'' (687299).



%

\bibliographystyle{IEEEtran}


\end{document}